\documentclass[12pt]{article}

\catcode`\@=11

\global\arraycolsep=2pt
\usepackage{amsbsy,amssymb,latexsym,amsfonts,amsmath}
\usepackage{graphicx,color}

\begin{document}
\begin{flushright}
\parbox{4.2cm}
{CALT 68-2933}
\end{flushright}

\vspace*{0.7cm}

\begin{center}
{ \Large Zoology of heterotic supercurrent supermultiplets in $d=2$}
\vspace*{1.5cm}\\
{Yu Nakayama}
\end{center}
\vspace*{1.0cm}
\begin{center}
{\it California Institute of Technology,  \\ 
452-48, Pasadena, California 91125, USA}
\vspace{3.8cm}
\end{center}

\begin{abstract}
We present various $(0,2)$ heterotic supercurrent supermultiplets in $(1+1)$ dimensional quantum field theories. From the minimal supercurrent supermultiplets, we deduce conditions on symmetry enhancement such as Lorentz invariance, (chiral) dilatation invariance, R-invariance, (chiral) conformal invariance and their various combinations. Our construction covers many interesting and/or exotic possibilities such as Lifshitz supersymmetry and warped superconformal algebra. We also discuss the corresponding supergravity by gauging the supercurrent supermultiplet. In particular, we propose a novel class of heterotic supergravity based on the virial supercurrent. 
\end{abstract}

\thispagestyle{empty} 

\setcounter{page}{0}

\newpage

\section{Introduction}
While the supersymmetry has been under thorough scrutiny for many decades, it is quite recent that the structure of the supercurrent supermultiplet has attracted a renewed attention \cite{Komargodski:2010rb}\cite{Dumitrescu:2011iu}. One problem often neglected in the earlier studies is that the supercurrent supermultiplet may not be unique and may admit a global structure, and the local nature of its existence in a previous formulation can affect its gauging, namely, the construction of the corresponding supergravity theories \cite{Komargodski:2009pc}\cite{Dienes:2009td}\cite{Kuzenko:2009ym}\cite{Butter:2010hk}\cite{Kuzenko:2010ni}\cite{Arnold:2012yi}. The detailed studies have cast subtle questions on the consistency of the Noether procedure. They have revealed that the minimal construction could become naive and it is sometimes more desirable to consider the globally well-defined supercurrent supermultiplet to make the consistency more transparent.

At the same time, reducibility conditions of the supercurrent supermultiplet uncover various hidden or manifest symmetries of the theory. For instance, the so-called R-multiplet makes the existence of the R-symmetry manifest \cite{Gates:1981yc}\cite{Gates:1983nr}. It is less familiar, but the existence of the so-called virial multiplet makes the existence of the dilatation symmetry manifest \cite{Kuzenko:2010am}\cite{Nakayama:2012nd}. The superconformal multiplet may be the most famous one, and the structure of the superconformal multiplet allows us to understand many important features of the superconformal field theories.\footnote{At this point, we would like to recall the distinction between dilatation invariance and the conformal invariance. We refer \cite{Nakayama:2013is} for a review of this subject.} 

Different supercurrent supermultiplets, in principle, imply different supergravity theories by gauging the corresponding distinct supercurrent. In practice, the difference typically lies only in auxiliary fields and the physics seems identical (at least classically). 
One famous example is the old minimal supergravity \cite{Stelle:1978ye}\cite{Ferrara:1978em}\cite{Fradkin:1978jq} based on the Ferrara-Zumino multiplet \cite{Ferrara:1974pz} and the new minimal supergravity \cite{Akulov:1976ck}\cite{Sohnius:1981tp} based on the R-multiplet. With the third possibility we just mentioned, it seems possible to construct the supergravity theories based on the virial multiplet. In $(1+3)$ dimension, although the corresponding linearized gravitational kinetic term is available \cite{Gates:2003cz}, the full non-linear theory has not been constructed so far.

In this paper, we address the similar issues in the $(0,2)$ heterotic supersymmetry in $(1+1)$ dimension. The structure of the supercurrent supermultiplet for the $(0,2)$ heterotic supersymmetry was studied in \cite{Dumitrescu:2011iu}, in which the manifest Lorentz invariance  was assumed. 
In our study, we do not explicitly assume the Lorentz invariance, and we treat the most generic structure of the supercurrent supermultiplet compatible with the $(0,2)$ heterotic supersymmetry. 
As we will see in section 2, there are various $(1+1)$ supersymmetric theories that have interesting symmetry structures with/without Lorentz invariance. For instance, our general construction accommodates the Lifshitz supersymmetry and supersymmetric generalization of the warped conformal algebra \cite{Hofman:2011zj}\cite{Detournay:2012pc}. Even with the Lorentz invariance, we propose the virial multiplet as a novel type of the reducibity condition not discussed in the literature, with which the dilatation invariance is made manifest.

Although we will see a plethora of supercurrent supermultiplets compatible with the $(0,2)$ heterotic supersymmetry with various symmetry enhancement, it is rather plausible that not all of them are realized in physically sensible situations. 
Indeed,  with some reasonably sounding assumptions, we can exclude some exotic possibilities due to the automatic symmetry enhancement. A canonical example of such is the enhancement of conformal invariance from the dilatation invariance and Lorentz invariance \cite{Zamolodchikov:1986gt}\cite{Polchinski:1987dy}. This was the rational to study conformal field theories in many physical applications. 
Similarly, we can formulate the enhancement of the chiral dilatation invariance to chiral conformal invariance as proposed in \cite{Hofman:2011zj}. It is an important question which supersymmetry multiplets are consistent with the symmetry enhancement argument. In section 3, we review the symmetry enhancement mechanism from our perspective and classify the compatible supersymmetric structures. 

In section 4, we discuss the gauging of the $(0,2)$ heterotic supercurrent supermultiplet with Lorentz invariance. As in $(1+3)$ dimension, a different choice of the supercurrent supermultiplet gives a different way to gauge the supercurrent. In $(1+1)$ dimension, the Einstein-Hilbert term is locally trivial, so the leading order supergravity only gives the constraint equations. We will see that  in addition to the known $(0,2)$ supergravity supermultiplets \cite{Brooks:1986uh}\cite{Evans:1986ada}\cite{Bergshoeff:1985gc}, the novel supergravity based on the virial multiplet is possible. The difficulty in $(1+3)$ dimension is avoided because there is no kinetic term. We propose an example of such theories by considering the improvement of the supercurrent supermultiplet of the heterotic sigma model.


\subsection{(0,2) supersymmetry and superspace}

The fundamental assumption of this paper is that the theory possesses the  $(0,2)$ supersymmetry that anti-commutes to be translation in one (light-cone) direction
\begin{align}
\{ Q_+ , \bar{Q}_+\} = -P_{++} \label{basis}
\end{align}
We also assume the existence of another (light-cone) translation $P_{--}$ that commute with $Q_+$, $\bar{Q}_+$ and $P_{++}$. 
In Lorentz invariant setup, we can regard $P_{\pm\pm}$ as the translation in light-cone directions $x^{\pm \pm}$ with the Minkowski metric $ds^2 = dx^{++} dx^{--}$, but our discussion in the following does not necessarily assume the {\it manifest} Lorentz invariance. We may regard $P_{++}$ as the Hamiltonian (time translation) and $P_{--}$ as momentum (space translation), or as in the opposite way, in non-relativistic setup.

To construct various supermultiplets, it is convenient to use the superspace formulation with the $(0,2)$ superspace $(x^{\pm}, \theta^+, \bar{\theta}^+)$. Our convention follows that of \cite{Dumitrescu:2011iu}, which in turn is related to the dimensional reduction of the the one in the textbook \cite{Wess:1992cp}.

In superspace, the supercharge $Q_{+}$ is represented by a differential operator as
\begin{align}
Q_{+} &= \frac{\partial}{\partial \theta^+} + \frac{i}{2} \bar{\theta}^+ \partial_{++} \cr
\bar{Q}_+ &= -\frac{\partial}{\partial \bar{\theta}^{+}} - \frac{i}{2} \theta^+ \partial_{++} 
\end{align}
acting on a superfield $\Phi(x^{++}, x^{--},\theta^+,\bar{\theta}^+)$. The superspace covariant derivative is defined as
\begin{align}
D_{+} &= \frac{\partial}{\partial \theta^+} - \frac{i}{2} \bar{\theta}^+ \partial_{++} \cr
\bar{D}_+ &= -\frac{\partial}{\partial \bar{\theta}^{+}} + \frac{i}{2} \theta^+ \partial_{++} 
\end{align}
with the anti-commutation relation
\begin{align}
\{D_+, \bar{D}_+ \} = i \partial_{++} \ .
\end{align}

The general superfield $\Phi(x^{++},x^{--},\theta^+,\bar{\theta}^+)$ may be reducible. We can introduce the purely algebraic constraint (in terms of the space-time derivatives) to define a chiral superfield
\begin{align}
\bar{D}_+ \Phi = 0 \cr
\end{align}
or an anti-chiral superfield
\begin{align}
D_+ \Phi^\dagger = 0
\end{align}
or a real superfield
\begin{align}
\Phi^\dagger = \Phi \ .
\end{align}
The  notation here is compatible with the rule $\bar{\theta}^+ = (\theta^+)^\dagger$.

\subsection{Noether assumption}
In this paper, we assume the existence of the conserved current for any symmetry generator. We call it the Noether assumption. This is not always necessary, nor is true because the axiom of local quantum field theories does not assume it, and we can find some counterexamples  (see e.g. \cite{Nakayama:2013is}:  generic higher spin theories do not possess the local energy-momentum tensor), but we restrict ourselves to the case in which the Noether assumption holds since otherwise the uplifting to the local symmetry (e.g. supergravity) seems extremely difficult. In particular, with the supersymmetry, the Noether assumption implies the existence of the supercurrent supermultiplet, and it is our main mission in this paper to study its structure. 

With our translation invariance \eqref{basis} and the Noether assumption, we require that for each symmetry of the theory there exists a conserved current, which satisfies
\begin{align}
\partial_{--} j_{++}^a + \partial_{++} j_{--}^a = 0 \ . \label{cons}
\end{align}
At this point, we note that there are several different ways to construct the ``conserved" charges out of the conserved current \eqref{cons} in our $(1+1)$ dimensional systems, depending on what we mean by the charge conservation. Probably, the most standard way to construct the conserved charge is based on the foliation generated by the ``time translation" $\partial_t \equiv \partial_{++} + \partial_{--}$. The conserved charge is given by
\begin{align}
Q^a = \int dx j_t^a \ ,
\end{align}
in which $x = x^{++} - x^{--}$, and $j_{t}^a = j_{++}^a + j_{--}^a$. 

In most of our discussions, we employ the light-cone structure and define
\begin{align}
Q^a = \int dx^{++} j^a_{++} + \int dx^{--} j^a_{--} \ 
\end{align}
which may be natural in ``radial quantization" of conformal field theories. Note that as far as the conservation concerns (i.e. $\frac{d}{dt} Q^a = 0$), the both definitions are fine.

As mentioned in section 1.1, the translation invariance and the corresponding conservation allow more generic interpretations than the interpretation based on the light-cone structure. After all, we have not assumed the Lorentz invariance so far, and the assumption of the light-cone structure may sound rather ad hoc.

One interesting class of such interpretations is to regard $P_{++}$ as Hamiltonian (time translation) and $P_{--}$ as momentum (space translation), or as in the opposite way, having non-relativistic applications in mind. 
The conservation equation \eqref{cons} is interpreted as 
\begin{align}
\partial_t j^a_t + \partial_x j^a_x = 0 
\end{align}
with the corresponding Noether charge $Q^a = \int dx j_t^a$. Note that the index interpretation here from the previous light-cone ones (e.g. $\partial_{++} \to \partial_t$ and $j_{--}^a \to j_t^a$) is {\it not} Lorentz covariant and it cannot be. Nevertheless it does not cause any difficulty. 
We will see the applications to the supersymmetric Lifshitz algebra in section 2.4.

The Noether assumption demands more than the existence of the conserved current. It requires that the constructed charges $Q^{a}$ generate the symmetry transformation
\begin{align}
i[Q^a, \mathcal{O} ] = \delta^a \mathcal{O} \label{symmetrytrans}
\end{align}
for any local (or non-local) Heisenberg operator $\mathcal{O}$. We should note that the conservation law alone does not fix the normalization of the Noether current, but the requirement of the symmetry transformation does fix the normalization of the Noether charge, and it is not allowed to change the normalization once it is fixed by \eqref{symmetrytrans}.\footnote{This is the origin of the non-renormalization of the charge operator. Note that this does not mean the non-renormalization of the Noether current itself because it can still be renormalized by adding the total derivative terms or trivially conserved terms without further assumptions.}

There are various different ways to realize the current conservation in $(0,2)$ superspace. One way is to introduce one chiral fermionic superfield $\bar{D}_+ \mathcal{J}^A_+ = 0$ and one anti-chiral left current superfield $J^A_{--}$ satisfying $D_+ J^A_{--} =0$. In components, we have
\begin{align}
\mathcal{J}^A_+ &= \lambda^A_+ + \theta^+ j^A_{++} - \theta^{+} \bar{\theta}^+ (\frac{i}{2}\partial_{++} \lambda_{+}^A) \cr
J_{--}^A &= j_{--}^A + \bar{\theta}^+ \psi_-^A + \theta^{+} \bar{\theta}^+ (\frac{i}{2}\partial_{++} j^A_{--}) \ .
\end{align}
The superspace current conservation reads:
\begin{align}
\partial_{--} \mathcal{J}^A_+ = i\bar{D}_+ J^A_{--} \ ,
\end{align}
which implies \eqref{cons} (and its complex conjugate).
The construction is minimal in the sense that there is no additional bosonic degrees of freedom than the current fields $j_{++}^A$ and $j_{--}^A$. Note, however, that since $(j_{++}^A, j_{--}^A)$ are complex fields, this multiplet contains the even numbers of conserved currents that are related by the supersymmetry. In $(0,2)$ superconformal setup, the multiplet is related to the $\mathcal{N}=2$ Kac-Moody current algebra \cite{Hull:1989py}.

Another possible conserved current multiplet is to demand 
\begin{align}
\bar{D}_+ (J^a_{--} + \frac{i}{2}\partial_{--} J^a) = 0 \ \label{realcurrent}
\end{align}
with a real superfield $J_{--}^a$ and $J^a$. In components
\begin{align}
J_{--}^a &= j^a_{--} + i\bar{\theta}^+ \bar{\lambda}_{+--}^a + i\theta^+ \lambda_{+--}^a + \theta^+ \bar{\theta}^+ K^a \cr
J^a &= L^a + i\bar{\theta}^+ \bar{\psi}^a_+ + i \theta^+ \psi^a_+ + \theta^+ \bar{\theta}^+ j^a_{++}
\end{align}
we have the conservation equation $\partial_{++} j_{--}^a + \partial_{--} j_{++}^a = 0$.

\section{Supercurrent supermultiplet}
Applying the Noether assumption to the $(0,2)$ supersymmetry algebra, we require the existence of a (not necessarily unique) energy-momentum tensor and a supercurrent:
\begin{align}
\partial_{--} T_{++++} + \partial_{++} T_{--++} &= 0 \cr
\partial_{--} T_{++--} + \partial_{++} T_{----} &= 0 \cr
\partial_{--} S_{+++}  - \partial_{++} \bar{s}_{--+} &= 0 \cr
\partial_{--} \bar{S}_{+++}  - \partial_{++} {s}_{--+} &= 0 \ .
\end{align}
We do not assume $T_{--++} = T_{++--}$ without the manifest Lorentz invariance.
Our first goal of this section is to uncover the superspace structure of this supercurrent conservation.

In superspace, the minimal assumption of the $(0,2)$ supersymmetry gives the following structure of the supercurrent supermultiplet
\begin{align}
\partial_{--} \mathcal{S}_{++} &= D_+ X_- - \bar{D}_+ \bar{X}_- \cr 
\bar{D}_+ \mathcal{T}_{----} &= \partial_{--} Y_{-} \ , \label{superc}
\end{align}
where $\mathcal{S}_{++}$ and $\mathcal{T}_{----}$ are real superfields. 
When the supersymmetry is not spontaneously broken,\footnote{See \cite{Dumitrescu:2011iu} for discussions on the broken case.} $X_-$ and $Y_{-}$ are both chiral $\bar{D}_+ X_- = \bar{D}_+ Y_{-} = 0$, and we can introduce (at least locally) the unconstrained potential superfield $\mathcal{X}_{--}$ and $\mathcal{Y}_{--}$ by $X_- = \bar{D}_+ \mathcal{X}_{--}$ and $Y_- = \bar{D}_+ \mathcal{Y}_{--}$ up to the gauge transformation $\mathcal{X}_{--} \to \mathcal{X}_{--} + \bar{D}_+ \mathcal{A}_{---}$ and $\mathcal{Y}_{--} \to \mathcal{Y}_{--} + \bar{D}_+ \mathcal{B}_{---}$. 
In components, we have
\begin{align}
\mathcal{S}_{++} &= j_{++} - i\theta^+ S_{+++} - i\bar{\theta}^+ \bar{S}_{+++} - \theta^+ \bar{\theta}^+ T_{++++} \cr
\mathcal{T}_{----} &= T_{----} -i\theta^+ \kappa_{---} -i\bar{\theta}^+ \bar{\kappa} _{---} + \theta^+ \bar{\theta}^+ T_{--} \cr
\mathcal{X}_{--} & = a_{--} + i\tilde{a}_{--} - \theta^+ \bar{{s}}'_{--+} - \bar{\theta}^+ {s}_{--+} + \theta^+ \bar{\theta}^+ (x+i\tilde{x}) \cr
X_- &= s_{--+} + \theta^+ \left( \frac{i}{2} \partial_{++} (a_{--} + i\tilde{a}_{--}) + (x+i\tilde{x}) \right) -\frac{i}{2} \theta^+ \bar{\theta}^+ \partial_{++} s_{--+} \cr 
\mathcal{Y}_{--} &= b_{--} + i\tilde{b}_{--} - \theta^+ \bar{t}'_{--+} - \bar{\theta}^+ t_{--+} + \theta^+ \bar{\theta}^+ (y+i\tilde{y}) \cr
Y_- &= t_{--+} + \theta^+ \left( \frac{i}{2} \partial_{++} (b_{--} + i\tilde{b}_{--}) + (y+i\tilde{y}) \right) -\frac{i}{2} \theta^+ \bar{\theta}^+ \partial_{++} t_{--+} \ .
\end{align}

The right conservation equation (the first line in \eqref{superc}) demands
\begin{align}
\partial_{--} j_{++} + \partial_{++} \tilde{a}_{--}  &= 2x \cr
\partial_{--} S_{+++} - \partial_{++} \bar{s}_{--+} & = 0 \cr
\partial_{--} T_{++++} +\partial_{++} T_{--++} &= 0 \cr
T_{--++} &= -\frac{\partial_{++} a_{--}}{2} - \tilde{x} \ , \label{rightcons}
\end{align}
and the left conservation equation (the second line in \eqref{superc}) demands
\begin{align}
i\bar{\kappa}_{---} &= \partial_{--} t_{--+} \cr
T_{--} &= -\frac{1}{2} \partial_{++} \partial_{--} \tilde{b}_{--} + \partial_{--} y \cr
\partial_{++} T_{----} + \partial_{--} T_{++--} &= 0 \cr
T_{++--} & = -\partial_{++}b_{--} -2\tilde{y} \ . \label{leftcons}
\end{align}
Thus, we can construct the conserved momenta
\begin{align}
P_{\pm\pm} = \int dx^{++} T_{++ \pm\pm} + \int dx^{--} T_{-- \pm \pm} \ .
\end{align}
 Note again that because we have not imposed the Lorentz invariance so far, the energy-momentum tensor is not necessarily symmetric $T_{++--} \neq T_{--++}$ at this point.

The $(0,2)$ supercharges can be constructed out of the conserved supercurrent as 
\begin{align}
Q_+ &= \int dx^{++} S_{+++} - \int dx^{--} \bar{s}_{--+} \cr
\bar{Q}_+ &= \int dx^{++} \bar{S}_{+++} - \int dx^{--} {s}_{--+} \ ,
\end{align}
and they satisfy the anti-commutation relation \eqref{basis}. Some useful supereymmetry transformations of component bosonic fields (note that $\delta_\epsilon = i \epsilon^+ Q_+ + i\bar{\epsilon}^+ \bar{Q}^+ = i\epsilon^+\frac{\partial}{\partial \theta^+} - i \bar{\epsilon}^+ \frac{\partial}{\partial \bar{\theta}^+} + \cdots$ )  are
\begin{align}
\delta_{\epsilon} T_{----} &= i\epsilon^{+} \partial_{--} \bar{t}_{+--} + i \bar{\epsilon}^+ \partial_{--} {t}_{+--} \cr
\delta_{\epsilon} j_{++} &=\epsilon^{+} S_{+++} - \bar{\epsilon}^+ \bar{S}_{+++} \cr
\delta_{\epsilon} T_{++++} &= -\frac{i}{2} \epsilon^+ \partial_{++} S_{+++} - \frac{i}{2}\bar{\epsilon}^+ \partial_{++} \bar{S}_{+++} \cr 
\delta_{\epsilon} (\partial_{++} a_{--} + 2\tilde{x}) &= - i\epsilon^+ \partial_{++} \bar{s}_{--+} - i \bar{\epsilon}^+ \partial_{++}s_{--+} \cr
\delta_{\epsilon} (\partial_{++} b_{--} + 2\tilde{y}) &= - i\epsilon^+ \partial_{++} \bar{t}_{--+} - i \bar{\epsilon}^+ \partial_{++}t_{--+} \cr
\delta_{\epsilon}(-\partial_{++} \tilde{a}_{--} + 2x) &= -\epsilon^+ \partial_{++} \bar{s}_{--+} + \bar{\epsilon}^+ \partial_{++} s_{--+} \ .
\end{align}

As usual, the supercurrent supermultiplet may  not be unique. The right conservation equation is invariant under
\begin{align}
\mathcal{S}_{++} &\to \mathcal{S}_{++} + [D_+,\bar{D}_+] U  \cr
X_- &\to X_- + \partial_{--} \bar{D}_+ U \ \label{rightimp}
\end{align}
for a real superfield $U$:
\begin{align}
U = u + i\theta^+ \alpha_+ + i\bar{\theta}^+ \bar{\alpha}_+ + \theta^+ \bar{\theta}^+ w_{++} \ .
\end{align}
It changes the energy-momentum tensor and the supercurrent as 
\begin{align}
\delta j_{++} & = 2w_{++} \cr
\delta T_{++++} & = \frac{1}{2}\partial^2_{++} u \cr
\delta S_{+++} & = i \partial_{++} \alpha_+ \cr
\delta s_{--+} &= -i \partial_{--} \bar{\alpha}_+ \cr
\delta x &= \partial_{--} w_{++} \cr
\delta a_{--} &= \partial_{--} u \cr
\delta T_{--++} &= -\frac{1}{2} \partial_{++} \partial_{--} u 
\end{align}
without affecting the conservation equation. 

Similarly the left conservation equation is invariant under
\begin{align}
\mathcal{T}_{----} \to \mathcal{T}_{----} + \partial_{--} V_{--} \cr
Y_{-} \to Y_- + \bar{D}_+ V_{--} \label{leftimp}
\end{align}
for a real superfield $V_{--}$:
\begin{align}
V_{--} = v_{--} + i\theta^+ \beta_{--+} + i\bar{\theta}^+ \bar{\beta}_{--+} + \theta^+ \bar{\theta}^+ z \ .
\end{align}
 It changes the energy-momentum tensor and the supercurrent as 
\begin{align}
\delta T_{----} &= \partial_{--} v_{--} \cr
\delta \kappa_{---} &= -\partial_{--} \beta_{--+} \cr
\delta T_{--} &= \partial_{--} z \cr
\delta t_{--+} &= -i\bar{\beta}_{--+} \cr
\delta  y &= z \cr
\delta b_{--} & = v_{--} \cr
\delta T_{++--} &= -\partial_{++} v_{--} 
\end{align}
without affecting the conservation equation.\footnote{Note again that the improvement here does not necessarily respect the Lorentz invariance.}

\subsection{Enhanced right symmetry}
It is possible that the right-moving supercurrent supermultiplet $\mathcal{S}_{++}$ and $\mathcal{X}_{--}$ are reducible within themselves. This, for instance, happens when the potential superfield $\mathcal{X}_{--}$ satisfies the reality condition
\begin{align}
\mathcal{X}_{--} = (\alpha + i\tilde{\alpha} ) \Xi_{--} \ 
\end{align}
with a real superfield $\Xi_{--}$.\footnote{Recall that the potential superfield $\mathcal{X}_{--}$ is defined only up to the gauge transformation $\mathcal{X}_{--} \to \mathcal{X}_{--} + \bar{D}_+ \mathcal{A}_{---}$, so strictly speaking, the reality condition is only up to the gauge transformation.}

At a generic value of the phase of $\alpha +i\tilde{\alpha} $, the supercurrent supermultiplet has an enhanced right ``dilatation" $D^R$ symmetry (when $\alpha \neq 0$)  under which
\begin{align}
i[D^R,P_{++}] = P_{++}  \ , \ \ i[D^R,P_{--}] = 0 \ .
\end{align}
To see this, we solve $\tilde{x} = \frac{\tilde{\alpha}}{\alpha} x$ from the reality condition and use the conservation equation \eqref{rightcons} to obtain
\begin{align}
T_{--++} = -\frac{\partial_{++} a_{--}}{2} - \frac{\tilde{\alpha}}{2\alpha} (\partial_{--} j_{++} + \partial_{++} \tilde{a}_{--}) \ . 
\end{align}
and we can construct the conserved right dilatation current
\begin{align}
D^R_{++} &= x^{++} {T}_{++++} + \frac{\tilde{\alpha}}{2\alpha} j_{++} \cr
D^R_{--} &= x^{++} T_{--++} + \frac{a_{--}}{2} + \frac{\tilde{\alpha}}{2\alpha} \tilde{a}_{--} \ \label{rightdilatation}
\end{align}
and the corresponding charge $D^R = \int dx^{++} D^R_{++} + \int dx^{--} D^R_{--}$.

It is interesting to observe that this right ``dilatation" acts on the supercharge as
\begin{align}
i[D^R, Q_{+} ] = (\frac{1}{2} - \frac{i\tilde{\alpha} }{2\alpha}) Q_{+} \ , \ \ 
i[D^R, \bar{Q}_{+} ] = (\frac{1}{2} + \frac{i\tilde{\alpha} }{2\alpha}) \bar{Q}_{+} 
\end{align}
and it has the twisted nature except $\tilde{\alpha} =0$ or $\alpha =0$. We call the supercurrent supermultiplet containing $\Xi_{--}$ as the virial multiplet when $\tilde{\alpha} =0$ because the existence of the virial multiplet guarantees the untwisted chiral right dilatation symmetry. Note that the virial multiplet does not necessarily imply the right special conformal invariance.

When $\alpha = 0$, the right dilatation current defined in \eqref{rightdilatation} is singular, and instead of the dilatation invariance, we have the R-symmetry. We call  the supercurrent supermultiplet containing $\Xi_{--}$  as R-multiplet when $\alpha=0$. We set $a_{--} =x=0$ in \eqref{rightcons} to obtain the conserved R-current 
\begin{align}
\partial_{--} j_{++} + \partial_{++} \tilde{a}_{--} = 0 
\end{align}
with the corresponding R-charge $R = \int dx^{++} j_{++} + \int dx^{--} \tilde{a}_{--}$ that generates the commutation relation
\begin{align}
i[R,Q_+] = -i Q_+ \ , \ \ i[R, \bar{Q}_+] =  i\bar{Q}_+ \ , \ \ [R,P_{++}] = [R,P_{--}] = 0  .
\end{align}
Therefore, the existence of the R-multiplet guarantees the $U(1)$ symmetry which is not associated with the space-time symmetry but acts on the supercharges.

If the theory possesses an extra conserved current, the R-current and virial current are not unique because we can add the conserved current to it without affecting the supermultiplet structure (unless we have extra assumptions such as superconformal invariance to enforce the preferred choice). For instance when we have a complex conserved current $\partial_{--} \mathcal{J}^A_+ = i\bar{D}_+ J^A_{--}$ with $\bar{D}_+ \mathcal{J}^A_+ = D_+ J^A_{--} = 0$ as discussed in section 1.2, one may improve the R-multiplet by
\begin{align}
\mathcal{S}_{++} &\to \mathcal{S}_{++} + D_+ \mathcal{J}^A_+ - \bar{D}_+ \bar{\mathcal{J}}^A_+ \cr
X_- &\to X_- + i\bar{D}_+ J^A_{--} \ .
\end{align}
Indeed, the conserved R-current $(j_{++},a_{--})$ is shifted by $(2\mathrm{Re} j_{++}^A$,$2\mathrm{Re} j_{--}^A)$ while the energy-momentum tensor $T_{\pm\pm ++}$ is shifted by $ \pm \partial_{\pm\pm} \mathrm{Im} j_{++}^A$. One can make the R-symmetry manifest by constructing pure imaginary $\mathcal{X}_{--}$ by rewriting $i\bar{D}_+ J^A_{--} = i\bar{D}_+(J^A_{--} +\bar{J}^A_{--})$ since $\bar{J}_{--}^A$ is chiral.

If, on the other hand, we have a real conserved current supermultiplet ($J^a$, $J_{--}^a$) introduced in section 1.2, we may improve the R-multiplet by
\begin{align}
\mathcal{S}_{++} &\to \mathcal{S}_{++} +  [D_+, \bar{D}_+] J^a \cr
X_- &\to X_- + \partial_{--} \bar{D}_+ J^a\cr
        &= X_- +2i \bar{D}_+ J_{--}^a \ .
\end{align}
In the second line, we used the current conservation \eqref{realcurrent} to make it manifest that the theory is still R-symmetric because $\mathcal{X}_{--}$ remains pure imaginary.

Similarly, we may improve the virial multiplet with
\begin{align}
\mathcal{S}_{++} &\to \mathcal{S}_{++} + D_+ \mathcal{J}^A_+ - \bar{D}_+ \bar{\mathcal{J}}^A_+ \cr
X_- &\to X_- + i\bar{D}_+ J^A_{--} \ .
\end{align}
The trace of the energy-momentum tensor $T_{--++}$ is shifted by $-\partial_{--} \mathrm{Im} j_{++}^A$. The dilatation invariance can be manifested by constructing real $\mathcal{X}_{--}$ by rewriting $i\bar{D}_+ J_{--}^A = i\bar{D}_+(J_{--}^A -\bar{J}_{--}^A)$ since $\bar{J}_{--}^A$ is chiral.

When $\mathcal{X}_{--}$ can be improved to be zero, the theory is right superconformal invariant. It is obvious from the above argument that it is both right dilatation invariant and R-invariant. This is consistent with the fact that we need the R-symmetry to close the $\mathcal{N}=2$ superconformal algebra.
Furthermore, the equations \eqref{rightcons} show $T_{--++} = 0$ and $s_{--+} = 0$, so we can construct the (infinitely) many conserved current as 
\begin{align}
J^{(n)} &= (x^{++})^n j_{++} \cr
G^{(n)} &= (x^{++})^n S_{+++} \cr
L^{(n)} &= (x^{++})^n T_{++++}  
\end{align}
 which generates  one copy of the familiar $\mathcal{N}=2$ super Virasoro algebra. 

One may wonder if it is possible to define the supercurrent for the theory which is R-invariant and right dilatation invariant but not right special conformal invariant. 
The theory must possess both the R-multiplet and the virial multiplet. At first sight, we might hasten to conclude that this is impossible because the potential $\mathcal{X}_{--}$ must be real (dilatation invariance), pure imaginary (R-invariance) and non-zero (not special conformal) simultaneously. A loophole here is that the supercurrent supermultiplet does not necessarily coincide for the virial multiplet and the R-multiplet. In addition, $\mathcal{X}_{--}$ has the gauge invariance, and the reality condition must hold only up to the gauge transformation.
The same theory may possess two (or more) distinct supercurrent supermultiplets, which are typically related by the improvement. 

For instance, when $\mathcal{X}_{--}$ is anti-chiral and it is given by $D_{+} \Upsilon_{---}$ with a complex fermionic superfield $\Upsilon_{---}$, then we can construct {\it both} the R-multiplet 
\begin{align}
\mathcal{X}_{--} = D_{+} \Upsilon_{---} + \bar{D}_+ \bar{\Upsilon}_{---}
\end{align}
and
the virial multiplet
\begin{align}
\mathcal{X}_{--} =  D_{+} \Upsilon_{---} - \bar{D}_+ \bar{\Upsilon}_{---}
\end{align}
 with the same $\mathcal{S}_{++}$ on the left hand side
\begin{align}
\partial_{--} \mathcal{S}_{++} = D_+ \bar{D}_+ \mathcal{X}_{--} - \bar{D}_+ D_+ \bar{\mathcal{X}}_{--} \ . \label{lefthandS}
\end{align}
The resulting theory is right dilatation invariant and R-invariant but not right special (super)conformal invariant (unless the right hand side of \eqref{lefthandS} vanishes up on improvement).
In more generic situations, the supercurrent supermultiplet $\mathcal{S}_{++}$ can be different  by certain improvement terms between the R-multiplet and the virial multiplet. Again we emphasize that the theory may not  necessarily be special (super)conformal invariant.

\subsection{Enhanced left symmetry}
Similarly, it is possible that the left supercurrent supermultiplet $\mathcal{T}_{----}$ and $\mathcal{Y}_{--}$ are reducible within themselves. It turns out that the reducibility condition is related to the left dilatation invariance.

The simplest case is when $\mathcal{Y}_{--}$ real. From \eqref{leftcons}, we have
\begin{align}
T_{++--} = -\partial_{++}b_{--}
\end{align}
This gives the left dilatation invariance and the left conformal invariance (as well as one copy of the enhanced Virasoro symmetry). To see the latter, we may define the improved energy-momentum tensor
\begin{align}
\tilde{T}_{----} &= T_{----} - \partial_{--} b_{--} \cr
\tilde{T}_{++--} &= T_{++--} + \partial_{++} b_{--} = 0 \ 
\end{align}
without spoiling the conservation. With this new ``traceless" energy-momentum tensor we can construct the left conserved Virasoro current
\begin{align}
\tilde{L}^{(n)} = (x^{--})^n \tilde{T}_{----} \ .
\end{align}

Probably a more direct way to realize this conformal improvement in a manifestly supersymmetric manner is to note that from \eqref{leftimp} one can make $\mathcal{Y}_{--}$  vanish by the improvement when it is real. Notice that the left conformal transformation (as well as the Virasoro extension) commutes with the supersymmetry $Q_{+}$ and $\bar{Q}_+$ so we do not necessarily require the superconformal invariance here.

Can we obtain left dilatation invariance without left special conformal invariance? One simple way to realize this option is as follows. Suppose $\mathcal{Y}_{--}$ is given by $\partial_{--} \Omega$ with a complex superfield $\Omega$ up to the improvement transformation by a real superfield $V_{--}$ in \eqref{leftimp}. It is left chiral scale invariant (but not necessarily chiral conformal invariant). To see this, we first note that the condition implies $\tilde{y} = \partial_{--} \zeta_{++}$, where $\zeta_{++}$ is the top component in $\Omega$, so that $T_{++--} = -\partial_{++} b_{--} - 2\partial_{--} \zeta_{++}$.
 Now it is possible to define the left dilatation  current
\begin{align}
D^{L}_{++} &= x^{--} T_{++--} + 2 \zeta_{++} \cr
D^{L}_{--} & = x^{--} T_{----} + b_{--} \ ,
\end{align}
which generates the left dilatation charge $D= \int dx^{++} D^{L}_{++} + \int dx^{--} D^{L}_{--}$ with the algebra
\begin{align}
i[D^L,P_{++}] = 0  \ , \ \ i[D^L,P_{--}] = P_{--}  \ , \ \ [D^{L}, Q_+] = [D^{L}, \bar{Q}_+] = 0 \ .
\end{align}
One may not be able to improve away $\zeta_{++}$, so the theory is not left special conformal invariant.
We will find a more elaborate way to obtain the left dilatation invariance without the left special conformal invariance in the following section.

The virial current supermultiplet is not unique when the theory possesses an extra conserved current. When we have a complex conserved current supermultiplet, we can improve
\begin{align}
\mathcal{T}_{----} &\to \mathcal{T}_{----} + \partial_{--} (J^A_{--} + \bar{J}^A_{--}) \cr
Y_{-} & \to Y_{-} + \bar{D}_+ (J^A_{--} + \bar{J}^A_{--}) \cr
 & = Y_{-} -i \partial_{--} \mathcal{J}^A_+ \ 
\end{align}
with noticing that $\mathcal{J}_+$ is chiral so we may introduce the potential $\mathcal{J}_+ = \bar{D}_+ \Omega$ and the left dilatation invariance is manifest from the above argument.

Similarly, when we have a real conserved current supermultiplet introduced in \eqref{realcurrent}, we can improve 
\begin{align}
\mathcal{T}_{----} &\to \mathcal{T}_{----} + \partial_{--} J^a_{--} \cr
Y_{-} &\to Y_{-} + \bar{D}_+ J^a_{--} \cr
      &=Y_{-} -\frac{i}{2} \bar{D}_+ \partial_{--} J^a  \ . 
\end{align}
The last expression makes the left dilatation manifest.

\subsection{Enhancement of left-right mixed symmetry}
So far, we discussed the reducibility of the supercurrent supermultiplet within  the right part or within the left part separately. Another important class of the reducibility comes from the relation between the right supercurrent supermultiplet and the left supercurrent supermultiplet. The most important condition is the left potential supermultiplet $\mathcal{X}_{--}$ and the right potential supermultiplet $\mathcal{Y}_{--}$ are proportional to each other: $\mathcal{Y}_{--} = (\beta + i\tilde{\beta}) \mathcal{X}_{--}$. The condition implies the existence of the extra enhanced ``dilatation" symmetry as we discuss in this section.

The parameter $\beta$ determines the dynamical critical exponent under which the dilatation acts on coordinate $x^{\pm\pm}$. On the other hand, the parameter $\tilde{\beta}$ determines how supercharge is twisted. $\beta = \tilde{\beta} =0$ is very special, and we will give a separate discussion in section 3 in our applications to the warped conformal field theories.

The particular case when $\tilde{\beta} = \frac{1}{2}$ is of great significance. This is because it implies the {\it manifest} Lorentz invariance. In \cite{Dumitrescu:2011iu}, the supercurrent supermultiplet with $\mathcal{Y}_{--} = \frac{1}{2}\mathcal{X}_{--}$ was named ``S-multiple".
 The conservation equations \eqref{leftcons} and \eqref{rightcons} demand that the energy-momentum tensor is symmetric:
\begin{align}
T_{++--} = T_{--++} 
\end{align}
so that the Lorentz current
\begin{align}
L_{++} = x^{--} T_{++--} - x^{++} T_{++++} \cr
L_{--} = x^{--} T_{----} - x^{++} T_{--++}  
\end{align}
is conserved. We can construct the Lorentz charge $ L = \int dx^{++} L_{++} + \int dx^{--} L_{--}$ that generates the Lorentz transformation.
 We can see that the supercharge transforms in the expected manner 
\begin{align}
i[L, Q_+] = -\frac{1}{2} Q_+ \ , \ \ i[L, \bar{Q}_+] = -\frac{1}{2} \bar{Q}_+ \ \ , \ \ i[L,P_{\pm\pm}] = \mp P_{\pm\pm} .
\end{align}
In particular, the supercharge is not twisted under the Lorentz transformation.
Therefore the system with the S-multiplet is manifestly invariant under the Poincar\'e $(0,2)$ supersymmetry.

As discussed in \cite{Dumitrescu:2011iu}, the S-multiplet has the improvement ambiguity under
\begin{align}
\mathcal{S}_{++} &\to \mathcal{S}_{++} + [D_+,\bar{D}_+] U  \cr
X_- &\to X_- + \partial_{--} \bar{D}_+ U \cr
\mathcal{T}_{----} &\to \mathcal{T}_{----} + \frac{1}{2}\partial^2_{--} U 
\end{align}
with a real superfield $U$. This is a particular combination of the improvement  \eqref{rightimp}\eqref{leftimp} which is consistent with the manifest Lorentz invariance.

The other extreme limit is $\beta = 0$ (say $\mathcal{Y}_{--} = i \mathcal{X}_{--}$), where the symmetry preserved by the condition is the left dilatation. 
\begin{align}
i[D^L,P_{++}] = 0  \ , \ \ i[D^L,P_{--}] = P_{--}  \ .
\end{align}
We can see it from the component expression for
\begin{align}
T_{++--} = -\partial_{++} b_{--} - 2 \tilde{y} = -\partial_{--}j_{++}
\end{align}
so that we can construct the conserved left dilatation current:
\begin{align}
D_{++}^L = x^{--} T_{++--} + j_{++} \ , \ \ D_{--}^L = x^{--} T_{----} \ .
\end{align}
At this point, we do not necessarily require the left special conformal transformation.
The left dilatation here is twisted in the sense that
\begin{align}
i[D^{L}, Q_+] = -iQ_+ \ , \ \ i[D^{L}, \bar{Q}_+] = i \bar{Q}_+ \ . 
\end{align}
More generic situations with a generic value of $\beta + i \tilde{\beta}$ will be studied further in relation to the Lifshitz (super)symmetry in section 2.4.

By combining the condition of the enhanced symmetries, we have a zoo of the heterotic supercurrent supermultiplets. We study some interesting examples with emphasis on the preserved Lorentz invariance in the rest of this section. We will see more exotic examples in later sections. 

(Example 1: manifest Lorentz invariance)

Let us first consider the {\it manifest} Lorentz invariant situations by assuming the existence of the S-multiplet with $\mathcal{Y}_{--} = \frac{1}{2}\mathcal{X}_{--}$. One may preserve the additional R-symmetry by assuming $\mathcal{X}_{--} = i\mathcal{R}_{--}$ is a pure imaginary superfield (R-multiplet). Or one may preserve the additional dilatation symmetry by assuming $\mathcal{X}_{--} = \mathcal{V}_{--}$ is a real superfield (virial multiplet). The dilatation invariance does not automatically imply the special (super)conformal invariance here. One may even retain the both R-invariance and dilatation invariance without preserve the special (super)conformal invariance if the theory admits both R-multiplet and virial multiplet. If we can improve the S-multiplet so that $\mathcal{Y}_{--} = \frac{1}{2}\mathcal{X}_{--} = 0$, the theory is conformal invariant with $(0,2)$ superconformal symmetry with one copy of super Virasoro extension.

(Example 2: left conformal, but not right superconformal)

It is one of special features of the two-dimensional space-time that the conformal algebra is reducible. We can preserve only the chiral half of the conformal symmetry without spoiling the Lorentz invariance and the dilatation invariance. Here, we show a possible supersymmetric extension of such an exotic situation. 

Again, let us first assume the manifest Lorentz invariance, so $\mathcal{Y}_{--} = \frac{1}{2}\mathcal{X}_{--}$. As discussed in section 2.1, we may be able to preserve the R-symmetry as well as the dilatation symmetry by further assuming $\mathcal{X}_{--}$ is anti-chiral so that $\mathcal{X}_{--} = D_{+} \Upsilon_{---}$. In components, we obtain the condition $x = -\frac{1}{2}\partial_{++} \tilde{a}_{--}$ and $\tilde{x} = \frac{1}{2} \partial_{++} a_{--}$ so that
\begin{align}
T_{++--} = T_{--++} = -\partial_{++} a_{--}
\end{align}
It is a divergence of a virial current, so the theory is dilatation invariant. R-current is also conserved
\begin{align}
\partial_{--} j_{++} + 2 \partial_{++} \tilde{a}_{--} = 0
\end{align}

The theory is not right superconformal invariant because there is no reason $X_{-}$ is improved away. However, it is actually left conformal invariant (with one copy of Virasoro extension). The point is that one may be able to improve the left energy-momentum tensor (at the sacrifice of the manifest Lorentz invariance) as
\begin{align}
\tilde{T}_{----} &= T_{----} - \partial_{--} a_{--} \cr
\tilde{T}_{++--} &= T_{++--} + \partial_{++} a_{--} = 0
\end{align}
so that we can construct the left Virasoro current
\begin{align}
\tilde{L}^{(n)} = (x^{--})^n \tilde{T}_{----} \ .
\end{align}
Note that we cannot do a similar improvement for $T_{--++}$ and $\tilde{T}_{++--} \neq T_{--++}$ indicates that the Lorentz invariance is not manifest (but it exists).

To make the discussion manifestly supersymmetric, 
we may construct the improved chiral left supercurrent supermultiplet as 
\begin{align}
\tilde{\mathcal{T}}_{----} &= \mathcal{T}_{----} -\frac{1}{2} \partial_{--} \left(D_+ \Upsilon_{---} - \bar{D}_+ \bar{\Upsilon}_{---} \right) \cr
\bar{D}_+ \tilde{\mathcal{T}}_{----} &=0 
\label{virrrr}
\end{align}
The manifest Lorentz invariance is lost here because $Y_- = \frac{1}{2}X_-$ is disguised with this left improvement.

The construction here gives an interesting spin-off. Suppose we begin with the right supercurrent supermultiplet which possesses both  R-symmetry and right dilatation symmetry by requiring $\mathcal{X}_{--} = D_{+} \Upsilon_{---}$. Instead of imposing the manifest Lorentz invariance, let us impose the left dilatation invariance  from $\mathcal{Y}_{--} = i\mathcal{X}_{--}$. The theory is manifestly right dilatation invariant and left Virasoro invariant because we may remove $\mathcal{Y}_{--}$ by the improvement:
\begin{align}
\tilde{\mathcal{T}}_{----} &= \mathcal{T}_{----} - i\partial_{--} \left(D_+ \Upsilon_{---} + \bar{D}_+ \bar{\Upsilon}_{---} \right) \cr
\bar{D}_+ \tilde{\mathcal{T}}_{----} &= 0 \ .
\end{align}
 Now, the theory must be Lorentz invariant because it is left dilatation invariant and right dilatation invariant. We can construct the manifestly Lorentz invariant supermultiplet by un-improving from \eqref{virrrr} as
\begin{align} 
\mathcal{T}^L_{----} &= \mathcal{T}_{----} - i\partial_{--} \left(D_+ \Upsilon_{---} + \bar{D}_+ \bar{\Upsilon}_{---} \right) +\frac{1}{2} \partial_{--} \left(D_+ \Upsilon_{---} - \bar{D}_+ \bar{\Upsilon}_{---} \right) \cr
\bar{D}_{+} \mathcal{T}^L_{----} &= \frac{1}{2}\partial_{--} \bar{D}_+ \mathcal{X}_{--} 
\end{align}
so that the Lorentz invariance is manifest. The twist in the right dilatation invariance is undone by the conserved R-current here.

\subsection{Lifshitz  supersymmetry}
Going back to the more generic values of $\beta + i\tilde{\beta}$ in the reducibility condition of the supercurrent supermultiplet  $\mathcal{Y}_{--} = (\beta + i\tilde{\beta}) \mathcal{X}_{--}$, the condition implies
\begin{align}
b_{--} &= \beta a_{--} - \tilde{\beta} \tilde{a}_{--} \cr
\tilde{b}_{--} & = \tilde{\beta} a_{--} + \beta \tilde{a}_{--} \cr
y &= \beta x - \tilde{\beta} \tilde{x} \cr
\tilde{y} &= \tilde{\beta} x + \beta \tilde{x} \ .
\end{align}
One may construct the conserved dilatation current
\begin{align}
D^{\beta,\tilde{\beta}}_{++} &= -2\beta x^{++} T_{++++} + x^{--} T_{++--} + \tilde{\beta} j_{++} \cr
D_{--}^{\beta,\tilde{\beta}} &= x^{--} T_{----} - 2\beta x^{++} T_{--++} 
\end{align}
so that the associated dilatation charge $D^{\beta,\tilde{\beta}} = \int dx^{++}D^{\beta,\tilde{\beta}}_{++}  + \int dx^{--} D^{\beta,\tilde{\beta}}_{--}$ satisfies the commutation relation
\begin{align}
i[D^{\beta,\tilde{\beta}}, P_{++}] = -2\beta P_{++} \ , \ \ i[D^{\beta,\tilde{\beta}} ,P_{--}]  = P_{--} \ .
\end{align}
Whenever $\tilde{\beta} $ is non-zero, the supersymmetry is twisted
\begin{align}
i[D^{\beta,\tilde{\beta}}, Q_+] &= (-\beta - {i}\tilde{\beta}) Q_+ \cr
i[D^{\beta,\tilde{\beta}}, \bar{Q}_+] &= (-\beta + {i} \tilde{\beta}) \bar{Q}_+ \ . 
\end{align}
If the theory has an extra R-symmetry, one may be able to undone the twist.

We can reinterpret the above result in terms of the Lifshitz supersymmetry. We regard either $P_{++}$ or $P_{--}$ as the Hamiltonian of the system, and $Q_+$ and $\bar{Q}_+$ as dynamical or kinematical supersymmetry respectively. The dynamical critical exponent is either $-2\beta$ or $-1/(2\beta)$. 

In the former situation, we identify $P_{++}$ as (minus of) the Hamiltonian $-H$ and $P_{--}$ as the momentum $P$. For a notational convention, we set $z = -2\beta$. The Lifshitz supersymmetry algebra is given by
\begin{align}
 [H,P] = \{Q,Q\} = \{\bar{Q},\bar{Q}\} = 0 \cr
i[D,H] = z H \ , \ \ i[D,P] = P \cr
\{Q,\bar{Q} \} = H \ .
\end{align}
The dynamical supersymmetry is twisted unless $\tilde{\beta} = 0$:
\begin{align}
i[D, Q] &= (\frac{z}{2} -{i}\tilde{\beta}) Q \cr
i[D, \bar{Q}] &= (\frac{z}{2} + {i} \tilde{\beta}) \bar{Q} \ . 
\end{align}
We may able to undone the twist if we have an extra R-symmetry.

As discussed in section 1.2, the light-cone conservation $\partial_{--} j_{++} + \partial_{++} j_{--} = 0$  is reinterpreted as $\partial_x j_{x} + \partial_t j_{t} = 0 $, and the corresponding charge is defined by $Q = \int dx j_t$. For instance, from the light-cone energy-momentum conservation
\begin{align}
\partial_{--} T_{++++} + \partial_{++} T_{--++} &= 0 \cr
\partial_{++} T_{----} + \partial_{--} \partial_{++--} &= 0
\end{align}
we have, being a little careful about tensor indices, the non-relativistic energy-momentum tensor conservation 
\begin{align}
\partial_{x} T_{x t} + \partial_{t} T_{t t}& = 0 \cr
\partial_{t} T_{tx} + \partial_x T_{xx} &= 0
\end{align}
so that $H =-\int dx T_{tt}$ and $P = \int dx T_{tx}$. 
The only subtle difference we may have to be careful is that when $j_{t} = j_{--} $ is zero, there is no corresponding charge, so the reducibility condition is slightly different.

The supercurrent supermultiplet and their structure can be easily recycled with  some minor changes in the tensor/spinor indices. The supercurrent conservation  can be represented as
\begin{align}
\partial_x \mathcal{S}_x &= D X - \bar{D} \bar{X} \cr
\bar{D} T_{tx} &= \partial_x Y 
\end{align}
with $D = \frac{\partial}{\partial \theta}-\frac{i}{2}\bar{\theta} \partial_t$.
The component expansion is
\begin{align}
\mathcal{S}_x &= j_x - i\theta S_x - i \bar{\theta} \bar{S}_x - \theta \bar{\theta} T_{xt} \cr
X & = s_t + \theta\left(\frac{i}{2}\partial_t (a_t + i\tilde{a}_t) + x + i\tilde{x} \right) - \frac{i}{2}\theta\bar{\theta} \partial_t s_t \cr
\mathcal{T}_{tx} &= T_{tx} - i\theta \kappa_{xt} - i \bar{\theta} \bar{\kappa}_{xt} + \theta \bar{\theta} T_x \cr
Y &= (\beta + i\tilde{\beta}) X \ .
\end{align}
The dynamical supercharge is realized as $Q=\int dx S_x$.

Going back to the other possibility, we have $z = -1/(2\beta)$ and the kinematical supersymmetry algebra is 
\begin{align}
 [H,P] = \{Q,Q\} = \{\bar{Q},\bar{Q}\} = 0 \cr
i[D,H] = z H \ , \ \ i[D,P] = P \cr
\{Q,\bar{Q} \} = -P \ .
\end{align}
The kinematical supersymmetry is twisted unless $\tilde{\beta} = 0$:
\begin{align}
i[D, Q] &= (\frac{1}{2} - {iz}\tilde{\beta}) Q \cr
i[D, \bar{Q}] &= (\frac{1}{2} + {iz} \tilde{\beta}) \bar{Q} \ . 
\end{align}
If we have an extra R-symmetry, we may undone the twist.
The supercurrent supermultiplet can be constructed in a similar manner essentially by exchanging $x$ with $t$:
\begin{align}
\partial_t \mathcal{S}_t &= D X - \bar{D} \bar{X} \cr
\bar{D} \mathcal{T}_{xt} &= \partial_t Y 
\end{align}
with $D = \frac{\partial}{\partial \theta}-\frac{i}{2}\bar{\theta} \partial_x$.
The component expression is 
\begin{align}
\mathcal{S}_t &= j_t - i\theta S_t - i \bar{\theta} \bar{S}_t - \theta \bar{\theta} T_{tx} \cr
X & = s_x + \theta\left(\frac{i}{2}\partial_x (a_x + i\tilde{a}_x) + x + i\tilde{x} \right) - \frac{i}{2}\theta\bar{\theta} \partial_x s_x \cr
\mathcal{T}_{xt} &= T_{xt} - i\theta \kappa_{tx} - i \bar{\theta} \bar{\kappa}_{tx} + \theta \bar{\theta} T_t \cr
Y &= (\beta + i\tilde{\beta}) X \ .
\end{align}
The kinematical supercharge is realized as $Q= \int dx s_x$. As mentioned earlier, the so-constructed dynamical or kinematical supercharges can be trivial when the original light-cone component satisfies $S_{+++} = 0$ or $s_{--+}=0$ respectively.

\section{Symmetry enhancement from unitarity condition}
The discussion in section 2 is based purely on the kinematics. The only non-trivial assumption is the Noether assumption that demands the existence of the supercurrent supermultiplet. This has enabled us to construct the supercurrent supermultiplet which does not respect the special (super)conformal invariance with the dilatation invariance. Although we have discussed possible symmetry enhancement from the structure of the supercurrent supermultiplet in section 2, the argument there does not directly imply that the symmetry must be enhanced.

Nevertheless, the additional physical as well as technical assumption makes the symmetry enhancement actually enforced. In this section, we discuss various conditions for the symmetry enhancement (e.g. chiral conformal invariance or full superconformal invariance) from unitarity, causality and so on.

The basic idea is based on the claim that chiral dilatation invariance must be enhanced to chiral conformal invariance as proposed by Hofman and Strominger \cite{Hofman:2011zj}. 
With this respect, it is important to realize that the chiral dilatation invariance (i.e. $[D^L,P_{++}] = 0$ or $[D^R,P_{--}]$ = 0) is special compared with the other non-chiral dilatation invariance as we will discuss. The Lorentz invariance is one example of non-chiral dilatation invariance, but it does not give any obvious symmetry enhancement.

The argument in \cite{Hofman:2011zj} goes as follows. Suppose the theory is chiral dilatation invariant. The Noether assumption dictates that the corresponding current exists. For a time being, we will focus on the left chiral dilatation since before we introduce the supersymmetry, which we will discuss in more details later, there is no essential difference between the left dilatation and right dilatation here.
With the translational invariance at hand, the chiral dilation invariance demands that the theory must possess the conserved energy-momentum tensor with the ``trace" given by the derivatives of a certain virial current $v_{\pm\pm}$
\begin{align}
\partial_{++} T_{----} + \partial_{--} T_{++--} &= 0 \cr
T_{++--} & = \partial_{++} v_{--} + \partial_{--} v_{++} \ 
\end{align}
so that the left chiral dilatation current 
\begin{align}
D_{++} &= x^{--} T_{++--} - v_{++} \cr
D_{--} &= x^{--} T_{----} - v_{--}
\end{align}
is conserved.
One may remove the left virial current $v_{--}$ on the right hand side by introducing the  improved energy-momentum tensor 
\begin{align}
\tilde{T}_{----} &= T_{----} + \partial_{--} v_{--} \cr
\tilde{T}_{++--} & = T_{++--} - \partial_{++} v_{--} = \partial_{--} v_{++} \ .
\end{align}
The corresponding charge is same as that of the original one up to total derivative contributions, and since we will assume that the chiral dilatation symmetry is not broken, the difference is unimportant.

Let us consider the two-point function $\langle v_{++}(x) v_{++} (0) \rangle$. The crucial assumption we employ here is that the (improved) energy-momentum tensor $\tilde{T}_{++--}$ has a canonical  chiral scaling dimension, and so does $v_{++}$. It was argued in \cite{Hofman:2011zj}\cite{Polchinski:1987dy}, this is possible whenever the theory is unitary and the dilatation operator is diagonalizable and the eigenvalues are discrete. If this is the case, we have $\langle v_{++}(x) v_{++} (0) \rangle =  f(x^{++})$ because the chiral scaling dimension of $\tilde{T}_{++--}$ is one and that of $v_{++}$ is zero. By taking the derivative twice, we obtain
$\langle \partial_{--} v_{++}(x) \partial_{--} v_{++}(0) \rangle = 0$ up to contact terms.\footnote{Certainly the condition is violated when the theory has a continuous spectrum. For instance, the target space rotation current for a free scalars $X^I \partial_{++} X^J - X^J \partial_{++} X^I$ has a non-holomorphic two-point function while the current is conserved.}

At this point, we need the additional assumption of the vacuum cyclicity. When any local operator $O(x)$ annihilates the vacuum,  $O(x) | 0 \rangle = 0$, then the operator itself vanishes $O(x) = 0$ identically. 
The Reeh-Schlieder theorem \cite{RS} shows that the statement is derivable in Lorentz invariant quantum field theories with microscopic causality, but we have not assumed the Lorentz invariance (manifestly), so this theorem is not necessarily justified \cite{Nakayama:2011fe}, but we will nevertheless assume it to draw the further conclusions.\footnote{Since we do not know any good examples beside the holographic construction, we do not know whether this is  reasonable or not \cite{Nakayama:2011fe}. Recently, a chiral Liouville action, which is non-local, non-compact, and non-Lagrangian, was proposed in \cite{Compere:2013aya}. See also other examples in \cite{Nakayama:2012ed}.} 
Together with the unitarity, the vacuum cyclicity demands $\partial_{--} v_{++} (x) = 0$, and therefore $\tilde{T}_{++--} = 0$ and the theory is right conformal invariant with the enhanced one copy of the Virasoro algebra generated by
\begin{align}
\tilde{L}^{(n)} = (x^{--})^{n} \tilde{T}_{----} \ .
\end{align} 

The similar argument applies to the right translation. The left chiral scale invariance and the assumption of the canonical chiral scaling dimension of the energy-momentum tensor demands  $\langle {T}_{++++}(x) T_{++++} (0) \rangle = g(x^{++})$ since the chiral scaling dimension of $T_{++++}$ is zero and therefore $\langle \partial_{--} {T}_{++++}(x) \partial_{--} T_{++++} (0) \rangle = 0$ up to possible contact terms. If we further assume the unitarity and the vacuum cyclicity, it means $\partial_{--} T_{++++} = \partial_{++} T_{--++} = 0$ from the conservation. When $T_{++++} \neq 0$, the theory is right conformal invariant (with Virasoro extension) generated by
\begin{align}
L^{(n)} = (x^{++})^{n} T_{++++} \ .
\end{align}
In this case, the theory is Lorentz invariant. On the other hand, when $T_{--++} \neq 0$, the theory has the left Kac-Moody current generated by
\begin{align}
K^{(n)} = (x^{--})^{(n)} T_{--++} \ .
\end{align} 
Note that when $T_{++++}=0$ 
the theory is not Lorentz invariant. In this particular case, the algebraic structure is known as the warped conformal algebra \cite{Hofman:2011zj}\cite{Detournay:2012pc}.

There are a couple of corollary with this argument.
One corollary (by applying the argument both in the left and right dilatation current simultaneously) is that when the theory is Lorentz invariant and dilatation invariant, then it must be (left and right) conformal invariant under the assumption of unitarity, causality, and the discreteness of the spectrum. 
This is nothing but the Zamolodchikov-Polchinski theorem \cite{Zamolodchikov:1986gt}\cite{Polchinski:1987dy} derived in a slightly different manner.

Another corollary is that the twisted supersymmetry algebra is always equivalent to the usual super(conformal) algebra unless the $(2,0)$ supersymmetry is further reducible. This does not mean that the twisted algebra is impossible, but it only means that the twisting can be undone without affecting the other conservations. We will see these examples below.

With the $(0,2)$ supersymmetry at hand, we have the following three types of the supercurrent supermultiplet that is consistent with the supersymmetric generalization of the Hofman-Strominger theorem. The situation depends on whether we start with the left or right chiral dilatation invariance. 

The rather trivial case is Lorentz invariant situation with the $(0,2)$ superconformal symmetry (with super Virasoro extension). It has the supercurrent supermultiplet with $X_{-} = Y_{-}= 0$:
\begin{align}
\partial_{--} \mathcal{S}_{++} &= 0 \cr
\bar{D}_+ \mathcal{T}_{----} &= 0 \ .
\end{align}
The super(conformal) current is chiral, and the left-mover and right-movers are decoupled. 

More non-trivial situations are related to  the warped (super)conformal algebra. We have two distinct possibilities where the supersymmetry is realized either in the same sector as the chiral conformal algebra or in the different sector.

Let us suppose the former case in which the supersymmetry is realized in the chiral conformal sector. The supercurrent supermultiplet must satisfy
\begin{align}
\partial_{--} \mathcal{S}_{++} &= 0 \cr
\partial_{--} Y_{-} & = 0 \cr
X_{-} & = \mathcal{T}_{-----} = 0 \ .
\end{align}
In components, we have the constraint $s_{--+} = a_{--} = \tilde{a}_{--} = x = \tilde{x} = T_{----} = T_{--} = \kappa_{---} = 0$, and the current algebra
\begin{align}
\partial_{--} j_{++} &= 0 \cr
\partial_{--} S_{+++} & = 0 \cr
\partial_{--} T_{++++}& = 0 \ ,
\end{align}
which generates the right $\mathcal{N}=2$ superconformal algebra (with super Virasoro extension). We also have the right $\mathcal{N}=2$ Kac-Moody current
\begin{align}
\partial_{--} T_{++--} &= \partial_{--} (-\partial_{++}b_{--} - 2\tilde{y}) = 0 \cr
\partial_{--} (\partial_{++} \tilde{b}_{--} -2 y) & = 0 \cr
\partial_{--} t_{--+} &= 0 \ ,
\end{align}
which generates the {\it left} translation (together with additional bosonic and fermionic current algebra extension) from $P_{--} = \int dx^{++} T_{++--}$. We note that the Kac-Moody algebra is naturally complexified from the $\mathcal{N}=2$ supersymmetry.

We may want to impose the additional constraint on $Y_-$ to reduce the supercurrent supermultiplet. Let us introduce the potential superfield $Y_- = \bar{D}_+ \mathcal{Y}_{--}$ and demand the reality condition on $\mathcal{Y}_{--}$ as discussed in section 2.2. It turns out that if you assume the unitarity, discreteness of the spectrum and the vacuum cyclicity, we conclude $Y_{-} = 0$ and  the theory is superconformal invariant (with trivial left moving sector).\footnote{The conclusion does not hold if the assumptions were violated and we could have a non-trivial real current superfield. Recall, however, we have already employed the assumption to restrict ourselves to the theories satisfying the warped conformal algebra, so there is less point in relaxing the condition here.}

Let us move on to the latter case in which the supersymmetry is realized in the  opposite sector to the chiral conformal algebra. 
 The supercurrent supermultiplet must satisfy
\begin{align}
D_+ X_- - \bar{D}_+ \bar{X}_- &= 0 \cr
\bar{D}_{+} \mathcal{T}_{----} &= 0 \cr
Y_{-} & = \mathcal{S}_{++} = 0 \ .
\end{align}
In components, we have the constraints $j_{++} = S_{+++} = T_{++++} = t_{--+} = b_{--} = \tilde{b}_{--} = y = \tilde{y} = 0$ and the current algebra
\begin{align}
\partial_{++} T_{----} &= 0 \cr
\kappa_{---} &= 0 \cr
T_{--} &= 0 \ ,
\end{align}
which generates the left conformal algebra (with Virasoro extension). 
Note that $T_{----}$ is singlet under the supersymmetry. We also have the left Kac-Moody current
\begin{align}
\partial_{++} \tilde{a}_{--} &= 2x \cr
\partial_{++} s_{--+} &=0 \cr
\partial_{++} T_{--++} &= \partial_{++}\left(-\frac{\partial_{++}a_{--}}{2} -\tilde{x}\right) = 0 \ ,
\end{align}
which generates the {\it right} supersymmetry algebra (together with bosonic and fermionic current algebra extension). 

We may want to impose the additional constraint on $X_-$ to reduce the supercurrent supermultiplet. Let us introduce the potential superfield $X_- = \bar{D}_+ \mathcal{X}_{--}$ and demand a reality condition on $\mathcal{X}_{--}$ as discussed in section 2.1. With a generic phase in the reality condition, the theory has the additional right chiral dilatation invariance, eventually it will lead to the full $(0,2)$ superconformal invariance with the trivial right sector $X_{-}=0$ \footnote{Therefore, the supersymmetry is trivial.} after imposing the unitarity, discreteness of the spectrum and the vacuum cyclicity. The only non-trivial case is to impose $\mathcal{X}_{--}$ to be pure imaginary. Then we have the extra R-current
\begin{align}
\partial_{++} \tilde{a}_{--} = 0 
\end{align}
as a part of the current algebra. We emphasize that we do not necessary have to require the R-symmetry for our $(0,2)$ supersymmetry with the warped conformal algebra.

\section{Gauging the heterotic supercurrent supermultiplet}
In this section, we would like to study possible gauging of the heterotic supercurrent supermultiplet.   Since our eventual interest is to construct the $(0,2)$ supergravity models, we will restrict ourselves to the case in which the Lorentz invariance is manifest. However, we note that our generic structure may be applied to the case without the manifest Lorentz invariance, and it would lead to non-relativistic supergravity such as foliation preserving diffeomorphic theories of gravity.

\subsection{Linearized analysis and super Virasoro constraint}
Let us consider the gauging of the S-multiplet.
\begin{align}
\partial_{--} \mathcal{S}_{++} &= D_+ X_- - \bar{D}_+ \bar{X}_- \cr
\bar{D}_+\mathcal{T}_{----} &= \frac{1}{2} \partial_{--} X_- \ .
\end{align}
We introduce  real superfields $H_{----}$, $H_{++}$ and a (fermionic) complex superfield $L_{-}$ with the linearized coupling
\begin{align}
S = \int d^2x d\theta^+ d\bar{\theta}^+ \left( H_{----} \mathcal{S}_{++} + L_{-} X_- - \bar{L}_- X_- + H_{++} \mathcal{T}_{----} \right) \ 
\end{align} 
as a superspace integral.\footnote{Our convention is $\int d\theta^+ d\bar{\theta}^+ (\theta^+ \bar{\theta}^+) = 1$.}

Because of the conservation equation, we can introduce the gauge symmetry to reduce the linearized heterotic supergravity multiplet
\begin{align}
H_{----} &\to H_{----} + \partial_{--} \Lambda_{--}  \cr
L_{-} &\to L_{-} - D_{+} \Lambda_{--} +\frac{1}{2}\partial_{--} M_+ + \bar{D}_+ N_{--} \cr
H_{++} &\to H_{++} + \bar{D}_+ {M}_+ - D_+ \bar{M}_+ \ , \label{ggauge}
\end{align}
where $N_{--}$ is an anti-chiral chiral superfield, $\Lambda_{--}$ is a real superfield and $M_+$ is a (fermionic) complex superfield. The gauge symmetry by $N_{--}$ for $L_{-}$ is because $X_-$ is a chiral superfield.

In components, the linearized heterotic  supergravity superfield can be expanded as 
\begin{align}
H_{----} & = h_{----} + i\theta^{+} \psi_{---} + i\bar{\theta}^{+} \bar{\psi}_{---} + \theta^+\bar{\theta}^+ A_{--} \cr
L_{-} & = \varphi_{-} + i\bar{\theta}^+ (h_{--++} + i B) + i\theta^+(\tilde{
h}_{--++} + iC) + \theta^+\bar{\theta}^+ \bar{\psi}_{++-} \cr
H_{++} & = 2h_{++} + i \theta^+ \sigma_{+++} + i\bar{\theta}^+ \bar{\sigma}_{+++} - \theta^+ \bar{\theta}^+ h_{++++} \ .
\end{align}
Let us take the minimal WZ gauge by using \eqref{ggauge}, so that $\tilde{h}_{--++} = C = \varphi_- = \sigma_{+++} = 0$. In this gauge, the linearized action becomes
\begin{align}
S = &\int d^2x \left( -h_{----} T_{++++} -2 h_{--++} T_{++--} - h_{++++} T_{----} \right. \cr
 &+  \bar{\psi}_{---} S_{+++} - \psi_{---} \bar{S}_{+++} + \bar{\psi}_{++-} s_{--+} - {\psi}_{++-} \bar{s}_{--+} \ \cr
 & \left. + A_{--} j_{++} + B \partial_{--} j_{++}  + h_{++} \partial^2_{--} j_{++} \right) \ .
\end{align}
The action is invariant under the linearized diffeomorphism\footnote{The reason why we have unfamiliar $\frac{1}{2}$ in the second line is due to the fact that the same component appears twice in the action due to the identification $h_{++--} = h_{--++}$.}
\begin{align}
\delta h_{----} &= \partial_{--} \xi_{--} \cr
\delta h_{--++} &= \frac{1}{2}\partial_{++} \xi_{--} + \frac{1}{2}\partial_{--} \xi_{++} \cr
\delta h_{++++} &= \partial_{++} \xi_{++} 
\end{align}
and the linearized superdiffeomorphism\footnote{Here the gauge symmetry by $M_{+}$ is used to preserve the Wess-Zumino gauge by compensating the fermionic transformation by $\Lambda_{--}$. This explains the apparent $\frac{1}{2}$ in the spin half gravitino transformation.}
\begin{align}
\delta \psi_{---} &= \partial_{--} \lambda_- \cr
\delta \bar{\psi}_{++-} &= \partial_{++} \lambda_-
\end{align}
as well as the compensated ``R-symmetry"
\begin{align}
\delta A_{--}  & = \partial_{--} \lambda  \cr
\delta B  & = \lambda + \partial_{--} \tilde{\lambda}_{++} \cr
\delta h_{++} &= -\tilde{\lambda}_{++} \ . 
\end{align}
We may further set $B= h_{++} = 0$ by using this Stuckerberg symmetry from the compensated R-gauging so that the effective degrees of freedom can be reduced to $(h, \psi, A_{--})$. In a sense, the R-symmetry is not gauged because we do not require that the matter is R-symmetric. It is rather compensated by auxiliary fields $(B,h_{++})$ in gravity multiplets.
The entire structure is consistent with the generic (ungauged) heterotic $(0,2)$ supergravity.

Let us move on to the gauging of the R-multiplet. The R-multiplet satisfies (renaming $\mathcal{R}_{++} = \mathcal{S}_{++}$, $\mathcal{R}_{--} = -i\mathcal{X}_{--}$ to emphasize the R-symmetry)
\begin{align}
\partial_{--} \mathcal{R}_{++} + \partial_{++} \mathcal{R}_{--} &= 0 \cr
\bar{D}_+ \left(\mathcal{T}_{----} - \frac{i}{2}\partial_{--} \mathcal{R}_{--} \right) &= 0 \ .
\end{align}
Since we have a well-defined potential superfield $\mathcal{R}_{--}$, we introduce a real superfield $H$ instead of the fermionic superfield $L_{-}$. The linearized coupling is
\begin{align}
S = \int d^2x d\theta^+ d\bar{\theta}^+ \left(H_{----} \mathcal{R}_{++} + H \mathcal{R}_{--} + H_{++} \mathcal{T}_{----} \right) \ .
\end{align}

Because of the conservation equation, we can introduce the gauge symmetry to reduce the linearized heterotic gauged supergravity multiplet
\begin{align}
H_{----} &\to H_{----} + \partial_{--} \Lambda_{--}  \cr
H &\to H + \partial_{++} \Lambda_{--} +\frac{i}{2}\partial_{--} (\bar{D}_+ {M}_+ + D_{+} \bar{M}_+) \cr
H_{++} &\to H_{++} +  \bar{D}_+ {M}_+ - D_+ \bar{M}_+ \ , \label{rgauge}
\end{align}
where $\Lambda_{--}$ is a real superfield and $M_+$ is a (fermionic) complex superfield defined up to the gauge transformation $M_+ \to M_+ + \bar{D}_+ \mathcal{M}$.

In components, the linearized heterotic gauged supergravity superfield can be expanded as
\begin{align}
H_{----} & = h_{----} + i\theta^{+} \psi_{---} + i\bar{\theta}^{+} \bar{\psi}_{---} + \theta^+\bar{\theta}^+ A_{--} \cr
H & = 2h_{--++} + i \theta^+ \bar{\psi}_{++-} + i\bar{\theta}^+ {\psi}_{++-} + \theta^+ \bar{\theta}^+ A_{++} \cr
H_{++} & = 2h_{++} + i \theta^+ \sigma_{+++} + i\bar{\theta}^+ \bar{\sigma}_{+++} - \theta^+ \bar{\theta}^+ h_{++++} \ .
\end{align}
We can take the Wess-Zumino gauge $\sigma_{+++} = h_{++} = 0$ from \eqref{rgauge}, and the linearized action becomes
\begin{align}
S = &\int d^2x \left(- h_{----} T_{++++} -2h_{--++} T_{++--} - h_{++++} T_{----} \right. \cr
 &+  \bar{\psi}_{---} S_{+++} - \psi_{---} \bar{S}_{+++} + \bar{\psi}_{++-} s_{--+} - {\psi}_{++-} \bar{s}_{--+} \ \cr
 & \left. + A_{--} j_{++} + A_{++} \tilde{a}_{--} \right) \ .
\end{align}
The action is invariant under linearized (super)diffeomorphism and the gauged R-symmetry. The latter is given by $A_{\pm\pm} \to A_{\pm\pm} = \partial_{\pm\pm} \lambda$ in line with the R-current conservation $\partial_{--} j_{++} + \partial_{++} \tilde{a}_{--} = 0$.

Finally, we study the gauging of the virial multiplet. The virial multiplet is given by (renaming  $\mathcal{V}_{++} = \mathcal{S}_{++}$, $\mathcal{V}_{--} = \mathcal{X}_{--}$ to emphasize the dilatation symmetry)
\begin{align}
\partial_{--} \mathcal{V}_{++} - [D_+,\bar{D}_+] \mathcal{V}_{--} &= 0 \cr
\bar{D}_+ \left(\mathcal{T}_{----} -\frac{1}{2}\partial_{--} \mathcal{V}_{--} \right) &= 0 \ .
\end{align}
Since we have a well-defined potential superfield $\mathcal{V}_{--}$, we introduce a real superfield $\tilde{H}$ instead of the fermionic superfield $L_{-}$. The linearized coupling is
\begin{align}
S = \int d^2x d\theta^+ d\bar{\theta}^+ \left(H_{----} \mathcal{V}_{++} + \tilde{H} \mathcal{V}_{--} + H_{++} \mathcal{T}_{----} \right) \ .
\end{align}

Because of the conservation equation, we can introduce the gauge symmetry to reduce the linearized heterotic virial supergravity multiplet
\begin{align}
H_{----} &\to H_{----} + \partial_{--} \Lambda_{--}  \cr
\tilde{H} &\to \tilde{H} +[D_+,\bar{D}_+] \Lambda_{--} +\frac{1}{2}\partial_{--} (\bar{D}_+ {M}_+ - D_{+} \bar{M}_+) \cr
H_{++} &\to H_{++} + \bar{D}_+ {M}_+ - D_+ \bar{M}_+ \ , \label{vgauge}
\end{align}
where $\Lambda_{--}$  is a real superfield and $M_+$ is a (fermionic) complex superfield defined up to the gauge transformation $M_+ \to M_+ + \bar{D}_+ \mathcal{M}$.

In components, the linearized heterotic virial supergravity superfield can be expanded as
\begin{align}
H_{----} & = h_{----} + i\theta^{+} \psi_{---} + i\bar{\theta}^{+} \bar{\psi}_{---} + \theta^+\bar{\theta}^+ A_{--} \cr
\tilde{H} & = 2h_{--++} + \theta^+ \bar{\psi}_{++-} - \bar{\theta}^+ {\psi}_{++-} + \theta^+ \bar{\theta}^+ A_{++} \cr
{H}_{++} & = h_{++} + i \theta^+ \sigma_{+++} + i\bar{\theta}^+ \bar{\sigma}_{+++} - \theta^+ \bar{\theta}^+ h_{++++} \ .
\end{align}
We can take the Wess-Zumino gauge $h_{--++} = \sigma_{+++} = h_{++}= 0 $ from \eqref{vgauge}, and the linearized action becomes
\begin{align}
S = &\int d^2x ( -h_{----} T_{++++} - h_{++++} T_{----} \cr
 & +  \bar{\psi}_{---} S_{+++} - \psi_{---} \bar{S}_{+++} + \bar{\psi}_{++-} s_{--+} - {\psi}_{++-} \bar{s}_{--+}  \cr
 &+ A_{--} j_{++} + A_{++} a_{--})  \ .
\end{align}
In this gauge, the fluctuation of the metric component is taken to be traceless.

The action is invariant under the linearized diffeomorphism (compensated by gauge transformation of the virial gauge fields $A_{++}$):
\begin{align}
\delta h_{----} &= \partial_{--} \xi_{--} \cr
\delta h_{--++} &= 0 \cr
\delta h_{++++} &= \partial_{++} \xi_{++} \cr
\delta A_{++} &= -\frac{1}{2}\partial_{++} (\partial_{++} \xi_{--} + \partial_{--} \xi_{++} ) \ .
\end{align}
Recall that $a_{--}$ is the virial current so that $T_{++--} = -\frac{\partial_{++} a_{--}}{2}$, and this ensures that the variation of the virial gauge field $A_{++}$ compensates the full diffeomorphism despite the absence of the $h_{--++}$ transformation.

To close this section, let us briefly discuss the supergravity equations of motions. All in the above constructions, the supergravity fields play the role of the Lagrange multiplier, and it enforces the supercurrent supermultiplet vanish.  In the S-multiplet case, we require $\mathcal{S}_{++} = \mathcal{T}_{----} = X_- = 0$. The R-multiplet case and the virial multiplet case are almost identical except that the condition on the R-current or virial current is slightly stronger. For comparison, suppose we consider the R-invariant or dilatation invariant matter. The S-multiplet gauging only requires vanishing of the derivative of the current e.g. $\partial_{++} a_{--} = 0$, while the R-multiplet or virial multiplet gauging require that the current itself vanishes (i.e. $a_{--} = 0$).

\subsection{Non-linear extension}
We would like to study some aspects of the non-linear (full supergravity) extension of the supercurrent gauging studied in the last section. While complete analysis with general matter couplings will be presented elsewhere, we give some salient features focusing on examples and general structures.

The first thing we note is that when we couple our different formulations of the supergravity to the superconformal matters, then the distinction essentially disappears after integrating out auxiliary fields and using the WZ gauge. To see this, let us consider the simplest example in which one free $(0,2)$ chiral multiplet couples with the $(0,2)$ heterotic supergravity in a superconformal manner.  The component action \cite{Brooks:1986uh}\cite{Evans:1986ada}\cite{Bergshoeff:1985gc} is given by
\begin{align}
\int d^2x \det(e_{m}^a) \left(-g^{mn}(\hat{D}_m \bar{X} \hat{D}_n X) -\frac{i}{2}[\bar{\chi}_+ (\hat{D}_{--}\chi_+) -(\hat{D}_{--} \bar{\chi}_+) \chi_+] +A_{--}\bar{\chi}_+ \chi_+ \right)  \label{sigmam}
\end{align}
with various covariant derivatives
\begin{align}
\hat{D}_{++} X &= \partial_{++} X \cr
\hat{D}_{--} X &= \partial_{--} X - \frac{i}{2} \psi_{---} \chi_+ \cr
\hat{D}_{--}\chi_+ &= (\partial_{--} + i\omega_{--}) \chi_+ -  (\partial_{++} X) \bar{\psi}_{---} \ . 
\end{align}
Here $e_m^a$ is the vielbein with the metric $g_{mn} = \eta_{ab} e_m^a e_n^b$, and $\omega_{---}$ is the spin connection. $X$ is a complex field and $\chi_+$ is a Weyl fermion. The index $\pm$ is the tangent (spinor) index, and $\partial_{\pm\pm} = e^{m}_{\pm\pm} \partial_m$. 
The spin $1/2$ component of the gravitino $\psi_{++-}$ as well as the Weyl mode of the metric decouples due to the superconformal invariance.

If we regard the action as the minimal supergravity corresponding to the gauging of the S-multiplet, we do not impose any gauge symmetry for the R-symmetry. $A_{--}$ is just the auxiliary field that will impose a part of the super Virasoro constraint. The action is still invariant under the super Weyl transformation (including the local R-symmetry), so we may want to gauge it eventually in applications to the $(0,2)$ critical string theory.

On the other hand, if we regard the action as the minimal R-gauged supergravity corresponding to the gauging of the R-multiplet, we have to impose the gauge transformation (as a part of the local supersymmetry)
\begin{align}
A_{--} &\to A_{--} + \partial_{--} \Lambda \cr
\chi &\to e^{i\Lambda} \chi \cr
\psi &\to e^{-i\Lambda} \psi \ .
\end{align}
The action \eqref{sigmam} is invariant under this local symmetry. We may further impose the super Weyl transformation, but in this case, the super Weyl transformation does not contain the R-symmetry. The resulting string model, nevertheless, is equivalent to the one from the S-multiplet gauging (because the R-symmetry gauging was already done).

Finally, if we regard the action as the minimal virial supergravity corresponding to the gauging of the virial multiplet, we have to impose the gauge transformation (as a part of the local supersymmetry)
\begin{align}
e_m^a &\to e^{\sigma} e_m^a \cr 
\chi &\to e^{-\frac{1}{2}\sigma} \chi \cr
\psi &\to e^{\frac{1}{2}\sigma} \psi \cr
A_{--} & \to A_{--} \cr
A_{++} & \to A_{++} + \partial_{++} \sigma \ .
\end{align}
Here $A_{++}$ is the virial connection that would cancel the non-invariance of the Weyl symmetry, but since the theory is superconformal, it did not appear in our action. This model still preserves the local (super) R-symmetry. In applications to the critical string theory, we may want to gauge it. After the gauging of the (super) R-symmetry, the resulting theory is the same as above two cases.


In order to make the distinctions more transparent, we would like to improve the energy-momentum tensor and break the superconformal invariance. We will consider the simplest improvement with $U = c\bar{X} X$ in \eqref{rightimp}. Such improvement should be regarded as the $(0,2)$ generalization of Fradkin-Tseytlin term \cite{Fradkin:1985ys}. The improvement we consider is consistent with S-multiplet, R-multiplet and virial multiplet. From the linearized coupling, we propose that the $(0,2)$ supergravity based on the S-multiplet acquires the extra terms:
\begin{align}
S_{\mathrm{S}} &= c\int d^2x  \det(e_{m}^a) \left( r  X\bar{X} - \left(\bar{X}\chi_+[(\partial_{--} + iA_{--})\psi_{++-}] + c.c \right) \right. \cr
 &  +  A_{--} (\bar{\chi}_+ \chi_+ + i (\bar{X} \partial_{++} X - X\partial_{++} \bar{X})) \cr
& \left. + ((\bar{X} \chi_+)D_{++} \psi_{---} + c.c) \right) , \label{my}
\end{align}
here $r$ is the two-dimensional curvature. 
The first line agrees with the proposal in \cite{Brooks:1987nt}, but we also need the remaining terms in order to explain our linearized supergravity. Our proposal agrees with the action based on the  full superspace integral proposed in \cite{Nibbelink:2012wb}. 

Note that in this minimal supergravity construction, $U$ can be an arbitrary function. 
The similar action based on the shift-symmetry of $X$ field rather than the phase rotation as the extra contribution to the R-symmetry was proposed in \cite{Berkovits:1993pk}, in which $U=c\bar{X}X$ is replaced by $U = (X+\bar{X})$. In their construction, the gauge fixing of $B$ field is done in a different manner.

Similarly, one can construct the $(0,2)$ supersymmetric Fradkin-Tseytlin term in the supergravity based on the R-multiplet because $U$ contains the conserved current. It is given by
\begin{align}
S_{\mathrm{R}} &= c\int d^2x  \det(e_{m}^a) \left( r  X\bar{X} - \left(\bar{X}\chi_+[(\partial_{--} + iA_{--})\psi_{++-}] + c.c \right) \right. \cr 
 &   +  A_{--} (\bar{\chi}_+ \chi_+ + i (\bar{X} \partial_{++} X - X\partial_{++} \bar{X})) - A_{++} (i (\bar{X} \partial_{--} X - X\partial_{--} \bar{X})) \cr
& \left. + ((\bar{X} \chi_+)D_{++} \psi_{---} + c.c) \right) - c^2\int d^2x \det(e_{m}^a) A_{++}A_{--} \bar{X}X \ . \label{myr}
\end{align}
 Note that in comparison with the S-multiplet gauging, the allowed $U$ is restricted from the necessity of the conservation.


Finally, let us elaborate on the virial current supergravity with the general construction.
We start with the theory with virial current $T_{++--} = \partial_{--} j_{++}$. We then introduce the compensating gauge field $A_{++} \to A_{++} + \partial_{++} \sigma$ under the Weyl transformation. Now since the non-Weyl invariance is the total derivative, we may construct the Weyl invariant theory with this compensating gauge field. To see the compatibility with the linearized analysis, we first set $\det g_{mn} = \mathrm{const}$. The diff $\times$ Weyl act as
\begin{align} 
\delta h_{----} &= \partial_{--} \xi_{--} \cr
\delta h_{--++} &= -\sigma + \partial_{++} \xi_{--} + \partial_{--} \xi_{++}  \cr
\delta h_{++++} &= \partial_{++} \xi_{++} \cr
\delta A_{++} &= \partial_{++} \sigma .
\end{align}
Now we set $\delta h_{--++} =0$ by choosing $\sigma = \partial_{++} \xi_{--} + \partial_{--} \xi_{++}$ and we recover our virial multiplet gauging. 

In our $(0,2)$ supersymmetric Fradkin-Tseytlin term, the above construction leads to the proposed action based on the virial supergravity:
\begin{align}
S_{\mathrm{V}} &= c\int d^2x  \det(e_{m}^a) \left( r  X\bar{X} - \left(\bar{X}\chi_+[(\partial_{--} + iA_{--})\psi_{++-}] + c.c \right) \right. \cr
 &  +  A_{--} (\bar{\chi}_+ \chi_+ + i (\bar{X} \partial_{++} X - X\partial_{++} \bar{X})) \cr
& \left. + ((\bar{X} \chi_+)D_{++} \psi_{---} + c.c) + A_{++} \partial_{--}(X\bar{X}) \right) \ . \label{myv}
\end{align}
At the classical level, more non-trivial improvement with general $U$ is possible. 
In all constructions, the supergravitational equations of motion gives the super Virasoro constraint we mentioned at the end of section 4.2.

To conclude this section, we realize that the would-be kinetic term (i.e. Einstein-Hilbert term) for the supergravity is a total derivative, and the corresponding supergravity extension, if any, is a total derivative. This is one distinction between our $d=2$ situation with the $d=4$ situation, in which the kinetic term for the virial supergravity seems more difficult to construct beyond the linearized approximation.

\section{Discussions}
In this paper, we have studied the zoology of $(0,2)$ heterotic supercurrent supermultiplets in $(1+1)$ dimension. We showed various possibilities in symmetry enhancement as reducibility conditions of the supercurrent supermultiplets. We identified the symmetry enhancement structure that is compatible with the unitarity and other technically assumptions.

We have show that there are three possible heterotic supercurrent supermultiplets that are consistent with the warped superconformal algebra. Concrete realizations of such supercurrent supermultiplets will be presented in a separate publication. These may be relevant for holographic constructions with supersymmetry.

In $(0,2)$ superconformal field theories, one important question we have to ask is which R-current and dilatation current are in the superconformal multiplet. This is practically very important because this will determine the operator spectrum, its chiral ring structure and the central charge. 

The former question was address in the recent papers \cite{Benini:2012cz}\cite{Benini:2013cda}\cite{Karndumri:2013iqa}. Whenever we have an additional chiral conserved current, it can mix with the superconformal R-current, and we have to identify the ``correct" superconformal R-current. The idea, similar to the $(1+3)$ dimensional problem \cite{Intriligator:2003jj}, is to start with the trial R-current
\begin{align}
j_{++}^{\mathrm{trial}} = j_{++}^{\mathrm{SCA}} + \sum_\alpha s_\alpha j_{++}^{\alpha} \ ,
\end{align}
where $\partial_{--} j_{++}^{\alpha} = 0$,
and compute the trial central charge $c_{\mathrm{trial}}$ from the 't Hooft anomaly matching condition, and finally maximize $c_{\mathrm{trial}}$. Under various assumptions such as unitarity, discreteness of the spectrum, and the absence of the accidental symmetry enhancement, the maximization condition leads to $s_{\alpha} = 0$. It will determine the ``correct" superconformal R-current if the so-determined R-charge assignment is physically sensible (e.g. non-violation of unitarity).\footnote{Otherwise, the theory may not be superconformal from the beginning. Or the accidental symmetry enhancement may occur.} We should note that this maximization does not necessarily assure that such a superconformal field theory exists.

We would like to ask a similar question for the (right) dilatation current. The situation is slightly more complicated because the trial dilatation current may contain the improvement terms:
\begin{align}
D_{++}^{R,\mathrm{trial}} &= x^{++}(T_{++++}^{\mathrm{SCA}} - \partial_{++}^2 u) -  \partial_{++}u +   \sum_\alpha \tilde{s}_\alpha j_{++}^{\alpha}  \cr
D_{--}^{R,\mathrm{trial}} &= x^{++} \partial_{--}\partial_{++} u  \ ,
\end{align}
where $\partial_{--} j_{++}^{\alpha} = 0$.
Since the presence of $u$ does not change the dilatation charge, we cannot determine $u$ from the requirement of the action of the dilatation on the other fields. 

One thing we do know from the unitarity is that there is no non-trivial dimension zero operator in compact conformal field theories, so unless $u$ is a trivial operator, the trial dilatation current does not have the canonical scaling dimension. In this way, we may select the purely right-moving dilatation current (i.e. $u=0$) as a part of the superconformal multiplet.

At this point, we are in the same position as in the R-multiplet. To go further, we have yet to determine $\tilde{s}_{\alpha}$ to fix the ``correct" superconformal dilatation current. We first note that the dilatation charge and other conserved flavor charges must commute from the conformal algebra, so they are simultaneously diagonalizable from the unitarity assumption. However, the unitary representation of the conformal algebra and the state operator correspondence demand that the eigenvalue of the ``correct" dilatation operator must be real while the action of the flavor symmetry must be pure imaginary. With this criterion, it is fairly easy to choose the ``correct" dilatation operator and the superconformal supercurrent supermultiplet.\footnote{Of course, as in the R-multiplet situation, we have tacitly assumed that the action of the symmetry on operators is known. It may not be the case in the strongly coupled situations, and then the discussion is powerless.}

The argument above can be applied to Lorentz invariant $(0,2)$ heterotic superconformal field theories, and it is not clear how much we can recycle these in the less symmetric situations like warped conformal field theories and their flows. A non-relativistic version of the $c$-theorem was analyzed from holographic viewpoints in \cite{Nakayama:2011fe}\cite{Myers:2010tj}, but the field theory counterpart is not yet established.

Our discussion may be useful to look for a non-trivial way to put the $(0,2)$ supersymmetric field theories on the curved manifold by utilizing the manifest dilatation symmetry. Another important issue for a future study is the higher dimensional supergravity based on the virial multiplet. We believe that the non-linear extension exists with the higher derivative kinetic terms (e.g. conformal supergravity), but whether the second order kinetic term is allowed or not is an open question.


\section*{Acknowledgements}
We would like to thank S.~J.~Gates, S.~Kuzenko, and S.~Groot Nibbelink for discussions.
This work is supported by Sherman Fairchild Senior Research Fellowship at California Institute of Technology  and DOE grant DE-FG02-92ER40701

\end{document}